\magnification=1200
\baselineskip=12pt 
\vsize=7.5in
\hsize=5in
\pageno=1
\null
\noindent
Lectures presented at the International School \hfil preprint SPhT/97-071\break
``Pair Correlation in Many-Fermion-Systems'',\hfil\break
Erice (Sicily), June 4 -- 10, 1997; to be published in \hfil\break
{\it Pair Correlation in Many-Fermion-Systems},\hfil\break
ed. V. Z. Kresin (Plenum, New York).\hfil\break
\vskip 1.0in
\noindent{\bf FULLERENE ANIONS AND PAIRING IN FINITE SYSTEMS}
\bigskip
\bigskip
\bigskip
\hskip 1.0in Th. Jolic\oe ur
\bigskip
\hskip 1.0in \vbox{\hbox{Service de Physique Th\'eorique}
\hbox{Orme des Merisiers}
\hbox{CEA Saclay}
\hbox{91191 Gif-s-Yvette, France}}

\bigskip
\bigskip
\bigskip
\noindent{\bf INTRODUCTION}
\medskip
The discovery of superconductivity in alkali-metal-doped fullerenes$^1$
K$_3$C$_{60}$ and Rb$_3$C$_{60}$ has raised interesting questions about
the electron-phonon coupling in such compounds and its interplay with
Coulomb repulsion. C$_{60}$ is a highly symmetrical molecule i.e. it is a
truncated icosahedron and its electronic lowest unoccupied molecular orbitals
(LUMO) are threefold degenerate$^{2,3,4}$. They form a T$_{1u}$ representation
of the icosahedral group {\bf I$_h$}.
Filling the LUMO in C$_{60}^{n-}$ anions
leads in a naive picture to narrow, partially filled bands in the
bulk fullerides. The bandwidth W is determined by the hopping between the
C$_{60}$ molecules which are quite far apart and W $\approx$ 0.5 eV. The
coupling of some H$_g$ phonons with electrons residing in the T$_{1u}$ orbital
has been suggested to be responsible for the superconductivity$^{5,6,7}$. The
Coulomb repulsion also may be important on the ball$^8$. Several
authors$^{9-12}$ have undertaken the study of the Jahn-Teller distortion that
 is
expected in the fullerene anions. In such calculations 
one approach is to consider the
electrons as fast degrees of freedom and the phonon normal coordinates are
treated as static$^{13}$. This is the strong-coupling approach
to the Jahn-Teller effect. Of course it is more
difficult to treat the fully dynamical problem of the phonon mode
coupled to the electrons.
Here we will investigate the interplay between the electronic and
phononic degrees of freedom on an {\it isolated} fullerene anion. 
We use a {\it weak coupling} approach to the Jahn-Teller effect
The lifting of degeneracy is obtained by a perturbation 
calculation in the case
of an undistorted anion. 
The ordering of levels can be described as "anti-Hund"' rule.
This calculation is very close in
spirit to the standard treatment of the electron-phonon coupling in
superconducting metals. In fact we show that this weak-coupling approach is exactly parallel to the nuclear physics calculation
of pairing of nucleons in a single shell. We recover the so-called
seniority model of pairing.
It is often claimed that ``superconductivity is a dynamical Jahn-Teller effect''. While such a statement is not very appropriate
to bulk metallic superconductors, we will show that in the context
of fullerene anions, it really makes sense.
There is now some evidence that the crossover from weak to strong coupling in this Jahn-Teller problem is smooth$^{24}$, so the weak-coupling
calculation captures the physics of the intermediate regime.

As in nuclear physics we find even-odd effects due to pairing
in the ions.
The effects we observe may
be sought by spectroscopy of solutions of fullerides in liquid ammonia,
for example. We discuss the opposite effect of Coulomb interaction, leading
to Hund's rule in ordinary situations. Finally we point out that experimentally
observed spectra may be at least partially explained by our  calculation$^{23}$.

\bigskip
\noindent{\bf THE ON-BALL ELECTRON-PHONON INTERACTION}
\medskip
The electronic structure of $\pi$ electrons in the C$_{60}$ molecule
is well known to be given by a simple H\"uckel calculation. The levels are
labeled$^4$ by the irreducible representations (irreps) of the icosahedron
group $\bf I_h$.  
One important property has to be noted: three of the $\bf I_h$ irreps are the
$l$=0,1,2 spherical harmonics of $\bf SO(3)$
 which do not split under the $\bf I_h$
group. They are commonly named A$_g$, T$_{1u}$, H$_g$. In
addition there is also the twofold spin degeneracy.

In the ground-state of the neutral C$_{60}$ molecule all levels up to
H$_u$ included are completely filled thus building a singlet state
$|\Psi_0\rangle$. The LUMO are the six T$_{1u}$ states.
These are occupied upon doping with extra electrons and the
ground-state becomes then degenerate. 
One then expects the Jahn-Teller effect to distort the anion and 
lift this orbital degeneracy$^{13,14}$.
due to the coupling of the T$_{1u}$ electrons to the vibrational modes of the molecule
(also referred to as phonons). 
In a weak-coupling scheme the phonon is purely virtual and is exchanged between electrons.
Phonon exchange between electrons leads to an
effective electron-electron interaction that competes with Coulomb repulsion
and may lead to anti-Hund ordering of energy-levels.

As a first investigation of electron-phonon coupling we use a perturbation
scheme suited to degenerate levels we will derive an effective
electron-electron interaction with the assumption that filled states lying
below the T$_{1u}$ level remain frozen so that intermediate states involve only
T$_{1u}$--T$_{1u}$ excitations. Indeed the H$_u$--T$_{1u}$ gap is 
$\approx$ 2eV whereas maximum phonon energies are $\approx$ 0.2eV.

A typical electron-phonon interaction term reads: 
$$ W=\sum_{\alpha
,m_1,m_2,\sigma}f_{\alpha m_1 m_2}^{}X_{\alpha}^{}
c_{m_1\sigma}^\dagger c_{m_2\sigma}^{} .$$
Here $X_\alpha$ are normal
coordinates, the subscript referring both to the irrep and to the row
in the irrep they belong to, $c_{m_1\sigma}^\dagger$ is the creation
operator for an electron with spin $\sigma$ in the T$_{1u}$ ($l$=1)
level, $m_1$ taking one of the $m$=$-$1,0,1 values, and $f_{\alpha m_1
m_2}$ are complex coefficients.  The $c_{m\sigma}^\dagger$ operators
transform as $l$=1 $|l,m\rangle$ vectors under $\bf I_h$ symmetries,
and their conjugates $c_{m\sigma}$ transform as $(-1)^{m+1}|l,-m\rangle$
vectors. The $(-1)^{m_2+1} c_{m_1\sigma}^\dagger c_{-m_2\sigma}^{}$ products
transform then as members of the T$_{1u}\times {\rm T}_{1u}$ representation,
which in the $\bf I_h$ group splits as: 
$$ {\rm T}_{1u}\times {\rm
T}_{1u}={\rm A}_g+{\rm T}_{1g}+{\rm H}_g .$$
This selects the possible
vibrational modes T$_{1u}$ electrons can couple to. 
In fact, only H$_g$ modes split the degeneracy$^{7}$.
     
Let us consider a particular fivefold degenerate multiplet of H$_g$
modes. Their normal coordinates will be labelled $X_m$, $m$ ranging
from -2 to +2. Since H$_g$ appears only once in the product
T$_{1u}\times {\rm T}_{1u}$, the interaction is determined up to one
coupling constant $g$ by the usual formula for the coupling of two
equal angular momenta to zero total angular momentum: 
$$ 
W=g\sum_m (-1)^m X_m \Phi_{-m} .
\eqno(1)
$$ 
The $X_m$ may be chosen such that $X_m^\dagger =(-1)^m X_{-m}^{}$
and have the following expression in terms of phonon operators: $$
X_m={1\over \sqrt 2}\left(a_m^{} +(-1)^ma_{-m}^\dagger \right)
\eqno(2)
$$ whereas the $\Phi_m$ are the irreducible $l$=2 tensor operators
built from the $c^\dagger c$ products according to: 
$$
\Phi_m^{} = \sum_{m_1} (1,1,2|m_1,m-m_1,m)(-1)^{(m-m_1+1)}
c_{m_1\sigma}^\dagger  c_{-m+m_1\sigma}^{} ,
\eqno(3)
$$ 
where $(l_1,l_2,l|m_1,m_2,m)$ are Clebsch-Gordan coefficients.

We now consider a doped C$_{60}^{n-}$, molecule, $0\le n \le6$. Its
unperturbed degenerate ground-states consist of $|\Psi_0\rangle$ to
which $n$ T$_{1u}$ electrons have been added times a zero-phonon
 state. They span a subspace denoted by ${\cal E}_0$.  In ${\cal
E}_0$ the unperturbed Hamiltonian $H_0$ reads: 
$$
H_0=\epsilon_{t_{1u}} \sum_{m,\sigma}c_{m\sigma}^\dagger
c_{m\sigma}^{} + \hbar\omega
\sum_m a_m^\dagger a_m^{},
$$ 
where $\epsilon_{t_{1u}}$ is the energy of the T$_{1u}$ level,
$\hbar\omega$ is the phonon energy of the H$_g$ multiplet under
consideration. Within ${\cal E}_0$ the effective Hamiltonian up to
second order perturbation theory is given by: 
$$
H_{eff}= E_0P_0 + P_0WP_0 + P_0W({\bf 1}-P_0){1\over {E_0-H_0}}({\bf
1}-P_0)WP_0 ,
$$ 
where $P_0$ is the projector onto ${\cal E}_0$, $E_0$
is the unperturbed energy in this subspace which is just the number of
doping electrons times $\epsilon_{t_{1u}}$. The linear term in $W$
gives no contribution. Using expressions (1) and (2) for $W$ and $X_m$
one finds: 
$$ 
H_{eff}=H_0 -{g^2\over {2\hbar\omega}}\sum_{m,\sigma_1
\sigma_2} (-1)^m \Phi_{m\sigma_1} \Phi_{-m\sigma_2},
\eqno(4)
$$ 
where we have now included spin indices. We can now use equation (3)
to express $H_{eff}$ as a function of $c$ and $c^\dagger$ operators
and put it in normal ordered form using fermion anticommutation rules.
In this process there appears a one-body interaction term which 
is a self-energy term.  We will henceforth omit the $H_0$ term which
is a constant at fixed number of doping electrons.

Let us now define pair creation operators
${A_{lm}^{s\sigma}}^\dagger$ which when operating on the vacuum
$|0\rangle$ create pair states of T$_{1u}$ electrons that are
eigenfunctions of $\bf L$,$\bf S$, $L_z$,$S_z$, where $\bf L$,$\bf S$
are total angular momentum and spin, and $L_z$,$S_z$ their
$z$-projections. $l$ and $s$ can take the values 0,1,2 and 0,1
respectively. This holds also if $|0\rangle$ is taken to be the
singlet state $|\Psi_0\rangle$.  
$$
{A_{lm}^{s\sigma}}^\dagger=\sum_{m_1,
\sigma_1}(1,1,l|m_1,m-m_1,m)({1\over 2},{1\over 2},s|
\sigma_1, \sigma-\sigma_1, \sigma)c_{m_1\sigma_1}^\dagger 
c_{m-m_1\sigma-\sigma_1}^\dagger .
\eqno(5)
$$ 
The quantity ${A_{lm}^{s\sigma}}^\dagger$ is non-zero only if $(l+s)$ is even
and the norm of ${A_{lm}^{s\sigma}}^\dagger|0\rangle$ is then equal
to $\sqrt 2$.  The inverse formula expressing $c^\dagger c^\dagger$ products
as $A^\dagger$ operators is: $$ c_{m_1\sigma_1}^\dagger
c_{m_2\sigma_2}^\dagger = \sum_{l,s}(1,1,l|m_1,m_2,m_1+m_2)
({1\over2},{1\over2},s|\sigma_1,\sigma_2,\sigma_1+\sigma_2) {A_{l
m_1+m_2}^{s \sigma_1+\sigma_2}}^\dagger .
\eqno(6)
$$ 
As $H_{eff}$ is a scalar, its two-body part may be written as a
linear combination of diagonal ${A_{lm}^{s\sigma}}^\dagger
{A_{lm}^{s\sigma}}^{}$ products whose coefficients depend only on $l$
and $s$: 
$$
\sum_{ls,m\sigma}F(l,s){A_{lm}^{s\sigma}}^\dagger A_{lm}^{s\sigma}.
$$ 
The $F(l,s)$ coefficients are calculated using expressions (4), (3), (6).
We then get $H_{eff}$ in final form: 
$$ 
H_{eff}=
-{5g^2\over{6\hbar\omega}} \left( \hat N + {A_{00}^{00}}^\dagger
A_{00}^{00}
-{1\over2}\sum_{m,\sigma}{A_{1m}^{1\sigma}}^\dagger A_{1m}^{1\sigma}
+{1\over10}\sum_m {A_{2m}^{00}}^\dagger A_{2m}^{00} \right) .
\eqno(7)
$$ 
In this formula $\hat N$ is the electron number operator for 
the T$_{1u}$ level;
the $\hat N$ term appears when bringing $H_{eff}$ of expression (4) in
normal ordered form. In our Hamiltonian formulation the
effective interaction is instantaneous.

There are actually eight H$_g$ multiplets in the vibrational spectrum
of the C$_{60}$ molecule.  To take all of them into account we only
have to add up their respective coefficients
$5g^2 / {6\hbar\omega}$, their sum will be called $\Delta$.

\bigskip
\noindent{\bf THE ELECTRONIC STATES OF FULLERENE ANIONS}
\medskip
We shall now, for each value of $n$ between 1 and 6, find the
$n$-particle states and diagonalize $H_{eff}$.  The Hamiltonian to be
diagonalized is that of equation (7) where the prefactor is replaced
by $-\Delta$. The invariance group of $H_{eff}$ is $\bf I_h \times
SU(2)$. The $n$-particle states may be chosen to be eigenstates of
$\bf L,S$, $L_z,S_z$ and we shall label the multiplets by $(l,s)$
couples, in standard spectroscopic notation ($^{2s+1}$L stands for (l,s)).
The pair $(l,s)$  label $\bf SO(3) \times SU(2)$ irreps which,
as previously mentioned, remain irreducible under $\bf I_h \times
SU(2)$ as long as $l$ doesn't exceed 2; for larger values of $l$ $\bf
SO(3)$ irreps split under $\bf I_h$. Fortunately enough, 
the relevant values of $l$ never exceed 2.  Moreover given any
value of $n$, $(l,s)$ multiplets appear at most once so that the
energies are straightforwardly found by taking the expectation value
of the Hamiltonian in one of the multiplet states. The degeneracies of
the levels will then be $(2l+1)(2s+1)$.  We now proceed to the
construction of the states.
\bigskip
$\bullet$ {\bf n=1:}
\noindent
There are six degenerate $^2$P states
$c_{m\sigma}^\dagger|\Psi_0\rangle$ whose energy is $-\Delta$.
\bigskip
$\bullet$ {\bf n=2:}
\noindent
There are 15 states, generated by applying
${A_{lm}^{s\sigma}}^\dagger$ operators on $|\Psi_0\rangle$. There is
one $^1$S state, five $^1$D states and  nine $^3$P states
with energies -4$\Delta$, -11$\Delta$/5, -$\Delta$.
\bigskip
$\bullet$ {\bf n=3:}
\noindent
There are 20 states.  States of given $l,m,s,\sigma$ can be built by
taking linear combinations of $A^\dagger c^\dagger |\Psi_0\rangle$
states according to: 
$$
\sum_{m_1,\sigma_1}(l_1,1,l|m_1,m-m_1,m)(s_1,{1\over2},
s|\sigma_1,\sigma-\sigma_1,\sigma)
{A_{l_1m_1}^{s_1\sigma_1}}^\dagger
c_{m-m_1\sigma-\sigma_1}^\dagger|\Psi_0\rangle .
$$ 
These states belong to the following multiplets: 
$^2$P (E=-3$\Delta$), $^2$D (E= -9$\Delta$/5), $^4$S (E=0).
\bigskip
$\bullet$ {\bf n=4:}
\noindent
There are 15 states, which are obtained by applying
${A_{lm}^{s\sigma}}^\dagger$ operators on
${A_{00}^{00}}^\dagger|\Psi_0\rangle$.
They are $^1$S (E=-4$\Delta$), $^1$D (E= -11$\Delta$/15),
$^3$P (E=-$\Delta$)
\bigskip
$\bullet$ {\bf n=5:}
\noindent
There are six $^2$P states which are $c_{m\sigma}^\dagger
{A_{00}^{00}}^\dagger {A_{00}^{00}}^\dagger|\Psi_0\rangle$ and whose
energy is $-\Delta$.
\bigskip
$\bullet$ {\bf n=6:}
\noindent
There is one $^1$S state whose energy is 0.

It is interesting to note that the above treatment of
electron-phonon interaction parallels that of pairing forces in atomic
nuclei$^{15,16}$. Of course in the case of finite fermionic systems there is no
breakdown of electron number but there are well-known "odd-even" effects
that appear in the spectrum. In our case pairing shows up in the $^1$S ground
state for C$_{60}^{2-}$ rather than $^3$P as would be preferred by
Coulomb repulsion i.e. Hund's rule. The construction of the states above is
that of the seniority scheme in nuclear physics$^{16}$.
We note that similar ideas have been put forward by V. Kresin some time ago,
also in a molecular context$^{17}$. The effective interaction that he
considered was induced by $\sigma$ core polarization.
\bigskip
\noindent{\bf THE EFFECT OF COULOMB REPULSION}
\medskip
We now consider the Coulomb 
electron--electron interaction and assume it to be small enough
so that it may be treated in perturbation theory. To get some feeling
of the order of magnitude of this repulsion we use the limiting
case of on-site interaction i.e. the Hubbard model. This Hamiltonian
is not specially realistic but should contain some of the Hund's rule physics.
The two-body interaction now reads: 
$$
{U\over2}\sum_{i,\sigma}c_{i\sigma}^\dagger c_{i-\sigma}^\dagger
c_{i-\sigma}^{}c_{i\sigma}^{} ,
$$
 where the $i$ subscript now labels the
$\pi$ orbitals on the C$_{60}$ molecule. 
The quantity U is $\approx$ 2-3 eV from quantum chemistry calculations$^{18}$
Since level degeneracies are split at first order
in perturbation theory we confine our calculation to this order
and have thus to diagonalize the perturbation	  
within the same subspace ${\cal E}_0$ as before.  In this subspace it
reads: 
$$ 
W_H =	U\sum_{i,\alpha\beta\gamma\delta}
\langle\alpha|i\rangle \langle\beta|i\rangle
\langle i|\gamma\rangle \langle i|\delta\rangle \ 
c_{\alpha\uparrow}^\dagger c_{\beta\downarrow}^\dagger
c_{\gamma\downarrow}^{} c_{\delta\uparrow}^{} ,
$$ 
where greek indices
label one--particle states belonging either to $|\Psi_0\rangle$ or to
the T$_{1u}$ level. Let us review the different parts of $W_H$. Note
that since the $|\Psi_0\rangle$ singlet remains frozen we have the identity:
$c_\alpha^\dagger c_\beta^{} = \delta_{\alpha\beta}$ if $\alpha,
\beta$ label states belonging to $|\Psi_0\rangle$.

\bigskip
--A part involving states belonging to $|\Psi_0\rangle$ only: $$
W_{H_1}=U \sum_{i, \alpha\beta} |\langle\alpha|i\rangle|^2
|\langle\beta|i\rangle|^2 \ c_{\alpha\uparrow}^\dagger
c_{\alpha\uparrow}^{} \ c_{\beta\downarrow}^\dagger
c_{\beta\downarrow}^{}
.
$$
$\alpha$,$\beta$ belong to
$|\Psi_0\rangle$. This term is thus diagonal within ${\cal E}_0$ and
merely shifts the total energy by a constant that does not depend on
the number of doping electrons. It won't be considered in the
following.
\bigskip
--A part involving both states belonging to $|\Psi_0\rangle$ 
and to the T$_{1u}$ level:
$$
W_{H_2}=U \sum_{i, \alpha\delta, \beta, \sigma} 
\langle\alpha|i\rangle \langle i|\delta\rangle |\langle\beta i|\rangle|^2 \
c_{\alpha\sigma}^\dagger c_{\delta\sigma}^{} \ 
c_{\beta-\sigma}^\dagger c_{\beta-\sigma}^{},
$$
where $\alpha,\delta$ belong to the T$_{1u}$ level whereas 
$\beta$ belongs to $|\Psi_0\rangle$. It reduces to:
$$
W_{H_2}=U\sum_{\alpha\delta, \sigma} c_{\alpha\sigma}^\dagger 
c_{\delta\sigma}^{} \ 
\bigg( \sum_i \langle\alpha|i\rangle \langle i|\delta\rangle 
\sum_\beta |\langle\beta|i\rangle|^2 .
\bigg)
$$
The sum over $\beta$ is just the density on site i for a given spin 
direction of all states
belonging to $|\Psi_0\rangle$ which is built out of completely filled
irreps. As a result this density is uniform and since $|\Psi_0\rangle$ 
contains 30 electrons for each spin
direction it is equal to 1/2. $W_{H_2}$ then becomes diagonal and reads:
$$
W_{H_2}= {U\over 2} \sum_{\alpha, \sigma} c_{\alpha\sigma}^\dagger 
c_{\alpha\sigma}^{} .
$$
Its contribution is thus proportional to the number of T$_{1u}$ 
electrons. It represents the interaction of
the latter with those of the singlet and we won't consider it in 
the following.
\bigskip
--A part involving only states belonging to the T$_{1u}$ level:

\noindent
$W_{H_3}$ has the same form as $W_H$ with all indices now belonging 
to the T$_{1u}$ level.
Whereas the interaction has a simple expression in the basis of
$|i\rangle$ states, we need its matrix elements in the basis of the 
T$_{1u}$ states.
There are in fact two T$_{1u}$ triplets in the one--particle spectrum 
of the C$_{60}$ molecule,
the one under consideration having higher energy.
To construct the latter we have first constructed two independent sets 
of states which transform
as $x,y,z$ under $\bf I_h$.
These are given by:
$$
|\alpha\rangle = \sum_i \vec {e_\alpha}.\vec {r_i} \ |i\rangle
\quad {\rm and} \quad 
|\alpha\rangle' = \sum_i \vec {e_\alpha}.\vec {k_i} \ |i\rangle ,
$$
where $\vec e_\alpha$ are three orthonormal vectors, $i$ labels sites on 
the molecule,
the $\vec r_i$ are the vectors joining the center of the molecule to the 
sites while the
$\vec k_i$ join the centre of the pentagonal face of the molecule the site 
$i$ belongs to
to the site i. We assume that the bonds all have the same length.
These states span the space of the two T$_{1u}$ triplets.
The diagonalization of the tight--binding Hamiltonian in the subspace of 
these six vectors 
yields then the right linear combination of the $|\alpha\rangle$ and 
$|\alpha\rangle'$ states
for the upper lying triplet.
From the $x,y,z$ states one constructs $l$=1 spherical harmonics.
We then get the matrix elements of $W_{H_3}$ in the basis of T$_{1u}$ states.
As ${\cal E}_0$ is invariant under $\bf I_h$ operations and spin rotations, 
$W_{H_3}$ which is the
restriction of $W_H$ to ${\cal E}_0$ is invariant too. It may thus be
expressed using the  $A$, $A^\dagger$ operators by using 
formula (5) in the same
way as the phonon--driven interaction and we finally get:
$$
W_{H_3}= \left( {U\over 40}{A_{00}^{00}}^\dagger A_{00}^{00} + {U\over 100}
\sum_m {A_{2m}^{00}}^\dagger A_{2m}^{00}\right) ,
\eqno(8)
$$
which is the only part in $W_H$ that we will keep.
Note that there is no contribution from $l$=1, $s$=1 $A^\dagger A$ products. 
Indeed the Hubbard
interaction is invariant under spin rotation and couples electrons having zero 
total $S_z$. As the coefficients
of $A^\dagger A$ products depend solely on $l$ and $s$ they must be zero for 
$s \ne 0$.
The spectrum for any number of T$_{1u}$ electrons is now easily found.
Of course the order of the multiplet is now reversed:
for n=2, we have  $^3$P, then $^1$D, then$^1$S. For n=3, we have
$^4$S, then $^2$D, then $^2$P. For n=4, we have $^3$P, then
$^1$D, then $^1$S.  
\bigskip
\vfill
\eject
\noindent{\bf CONCLUSION}
\medskip
The ordering of energy levels in the electron-phonon scheme are clearly
opposite to those of Hund's rule. 
The clear signature
of what we can call "on-ball" pairing is the ground state $^1$S of 
$C^{2-}_{60}$: the two extra electrons are paired by the electron-phonon
coupling. We note that the U of the Hubbard model appears divided by large
factors: this is simply due to the fact that the C$_{60}$ molecule is large.
As a consequence, if U $\approx$ 2 eV, Coulomb repulsion may be overwhelmed
by phonon exchange. With a H$_g$ phonon of typical energy 100 meV and
coupling O(1) as suggested by numerous calculations$^{6,7,10}$, the quantity
$\Delta$ may be tens of meV. 

It seems to us that the cleanest way to probe
this intramolecular pairing would be to look at solutions of fullerides leading
to free anions such as liquid ammonia solutions or organic solvents$^{19-22}$. 
EPR or IR
spectroscopy should be able to discriminate between the two types of spectra.
Measurements by EPR should determine whether or not the two extra electrons
in C$^{2-}_{60}$ are paired, for example. In near-IR spectroscopy the lowest
allowed transition for C$^{2-}_{60}$ should be at higher energy than that of
C$^{-}_{60}$ due to the pairing energy
while in the Coulomb-Hubbard case it is at lower energy.

Present experiments$^{19,20}$ have studied the near-IR spectra of solutions of
fulleride anions prepared by electrochemical reduction.
There are several peaks that do not fit
a simple H\"uckel scheme of levels. They do not have an immediate 
interpretation in terms of vibrational structure$^{19,20}$.
With our energy levels in table I, a
tentative fit would lead to $\Delta \approx$ 80 meV assuming U = 0.
Such a value leads to intriguing agreement with the major peaks seen
for C$^{2-}_{60}$ and C$^{3-}_{60}$ while this is no longer the case for
C$^{4-}_{60}$ and C$^{5-}_{60}$.

Finally we mention that recent EPR experiments$^{22}$ have given some
evidence for non-Hund behaviour of the fulleride anions.
While one may observe
some trends similar to the results of the phonon-exchange approximation,
it is clear that the model we used is very crude. In a bulk {\it conducting} solid we do not
expect the previous scheme to be valid since the levels are broadened into
bands: then phonon exchange leads of course to {\it superconductivity}.

\bigskip
\noindent
{\bf Acknowledgments}
\bigskip
This work has been done in collaboration with Lorenzo Bergomi
and is a part of his PhD thesis.
We thank K. M. Kadish, M. T. Jones, C. A. Reed and J. W. White for informing us
about their recent experimental work. We have
also benefited from the help of the GDR "Fullerenes" supported by the Centre
National de la Recherche Scientifique (France).
Thanks are also due to V. Z. Kresin for his kind invitation 
to present this work in Erice summer school.
\vfill
\eject
\centerline{\bf REFERENCES}
\bigskip
\item{[1]}For a review see A. F. Hebard, Physics Today, November 1992.
\medskip
\item{[2]} S. Satpathy, Chem. Phys. Letters {\bf 130}, 545 (1986).
\medskip
\item{[3]} R. C. Haddon and L. T. Scott, Pure Appl. Chem. {\bf 58}, 
137 (1986); R. C. Haddon, L. E. Brus, and K. Raghavachari, 
Chem. Phys. Letters {\bf 125}, 459 (1986).
\medskip
\item{[4]} G. Dresselhaus, M. S. Dresselhaus, and P. C. Eklund, Phys. 
Rev. B{\bf 45}, 6923 (1992).
\medskip
\item{[5]}M. Lannoo, G. A. Baraff, M. Schl\"uter and D. Tomanek, Phys. Rev.
B{\bf 44}, 12106 (1991). See also K. Yabana and G. Bertsch, Phys. Rev. B{\bf
46}, 14263 (1992).
\medskip
\item{[6]} V. de Coulon, J. L. Martins, and F. Reuse, 
Phys. Rev. B{\bf 45}, 13671 (1992).
\medskip
\item{[7]} C. M. Varma, J. Zaanen, and K. Raghavachari, 
Science {\bf 254}, 989 (1991).
\medskip
\item{[8]} S. Chakravarty and S. Kivelson, Europhys. Lett.
{\bf 16},  751 (1991); Chakravarty, M. P. Gelfand, and S. Kivelson, Science 
{\bf 254}, 970 (1991); S. Chakravarty, S. Kivelson, M. Salkola, and S. Tewari, 
Science {\bf 256}, 1306 (1992).
\medskip
\item{[9]}A. Auerbach, Phys. Rev. Lett. {\bf 72}, 2931 (1994).
\medskip
\item{[10]}N. Koga and K. Morokuma, Chem. Phys. 
Letters {\bf 196}, 191 (1992).
\medskip
\item{[11]} J. C. R. Faulhaber, D. Y. K. Ko, and P. R. 
Briddon, Phys. Rev. B{\bf 48}, 661 (1993).
\medskip
\item{[12]}F. Negri, G. Orlandi and F. Zerbetto, Chem. Phys. Lett. {\bf 144},
31 (1988).
\medskip
\item{[13]}M. C. M. O'Brien and C. C. Chancey, Am. J. Phys. {\bf 61}, 688
(1993).
\medskip
\item{[14]}J. B. Bersuker, The Jahn-Teller Effect 
and Vibronic Interactions in Modern Chemistry, 
Plenum Press, (1984).
\medskip
\item{[15]}P. Ring and P. Schuck, The Nuclear Many-Body Problem, 
Springer-Verlag, (1980).
\medskip
\item{[16]}J. M. Eisenberg and W. Greiner, "Nuclear Theory, 
Microscopic Theory of the Nucleus", North Holland, Amsterdam (1972), Vol. III, see pp. 287-317.
\medskip
\item{[17]}V. Z. Kresin, J. Supercond. {\bf 5}, 297 (1992);
V. Z. Kresin, V. A. Litovchenko and A. G. Panasenko, J. Chem. Phys. 
{\bf 63}, 3613 (1975). 
\medskip
\item{[18]}V. P. Antropov, O. Gunnarsson and O. Jepsen, Phys. Rev. B{\bf 46},
13647 (1992).
\medskip
\item{[19]} G. A. Heath, J. E. McGrady, and R. L. Martin, 
J. Chem. Soc., Chem. Commun. 1272 (1992).
\medskip
\item{[20]} W. K. Fullagar, I. R. Gentle, G. A. Heath, and 
J. W. White, J. Chem. Soc., Chem. Commun. 525 (1993).
\medskip
\item{[21]}D. Dubois, K. M. Kadish, S. Flanagan and L. J. Wilson, J. Am. Chem.
Soc. {\bf 113}, 7773 (1991); ibid. {\bf 113}, 4364 (1991).
\medskip
\item{[22]}P. Bhyrappa, P. Paul, J. Stinchcombe, P. D. W. Boyd and C. A. Reed,
"Synthesis and Electronic Characterization of Discrete Buckminsterfulleride
salts", JACS, November 1993.
\medskip
\item{[23]}L. Bergomi and Th. Jolicoeur, Compte-Rendus de l'Acad\'emie des Sciences (Paris), {\bf 318} II, 283 (1994),
cond-mat/9311011.
\medskip
\item{[24]}A. Auerbach, N. Manini and E. Tosatti, Phys. Rev. 
B{\bf 49}, 12998 (1994).
\vfill
\eject
\bye